\documentclass[conference]{IEEEtran}
\usepackage{amssymb}
\usepackage{times,amsmath,epsfig,float}
\usepackage{subfigure}
\usepackage{graphicx}
\title{Adaptive Distributed Space-Time Coding in Cooperative MIMO Relaying Systems using Limited Feedback}

\author{{Tong Peng $\S$, Rodrigo C. de Lamare $\S$ and Anke Schmeink
$\clubsuit$}\\
$\S$ Communications Research Group, Department of Electronics, University of York, York YO10 5DD, UK\\
$\clubsuit$ UMIC Research Centre, RWTH Aachen University, D-52056
Aachen, Germany \\
Emails: tp525@ohm.york.ac.uk; rcdl500@ohm.york.ac.uk,
schmeink@umic.rwth-aachen.de}

\begin{document}


\maketitle
\IEEEpeerreviewmaketitle

\begin{abstract}
An adaptive randomized distributed space-time coding (DSTC) scheme
is proposed for two-hop cooperative MIMO networks. Linear minimum
mean square error (MMSE) receiver filters and randomized matrices
subject to a power constraint are considered with an
amplify-and-forward (AF) cooperation strategy. In the proposed DSTC
scheme, a randomized matrix obtained by a feedback channel is
employed to transform the space-time coded matrix at the relay node.
The effect of the limited feedback and feedback errors are
considered. Linear MMSE expressions are devised to compute the
parameters of the adaptive randomized matrix and the linear receive
filters. A stochastic gradient algorithm is also developed with
reduced computational complexity. The simulation results show that
the proposed algorithms obtain significant performance gains as
compared to existing DSTC schemes.
\end{abstract}

\section{Introduction}

Cooperative multiple-input and multiple-output (MIMO) systems, which
employ multiple relay nodes with antennas between the source node
and the destination node as a distributed antenna array, apply
distributed diversity gain and provide copies of the transmitted
signals to improve the reliability of wireless communication systems
\cite{Clarke}. Among the links between the relay nodes and the
destination node, cooperation strategies, such as
Amplify-and-Forward (AF), Decode-and-Forward (DF), and
Compress-and-Forward (CF) \cite{J. N. Laneman2004} and various
distributed space-time coding (DSTC) schemes in \cite{J. N.
Laneman2003}, \cite{Yiu S.} and \cite{RC De Lamare} can be employed.

The utilization of a distributed STC (DSTC) at the relay node in a
cooperative network, providing more copies of the desired symbols at
the destination node, can offer the system diversity gains and
coding gains to combat the interference. The recent focus on the
DSTC technique lies in the design of delay-tolerant codes and
full-diversity schemes with minimum outage probability. An
opportunistic DSTC scheme with the minimum outage probability is
designed for a DF cooperative network and compared with the fixed
DSTC schemes in \cite{Yulong}. An adaptive distributed-Alamouti
(D-Alamouti) STBC design is proposed in \cite{Abouei} for
non-regenerative dual-hop wireless systems which achieves the
minimum outage probability.

The channel state information (CSI) is very important for a wireless
communication system and can be estimated by sending a block of
training symbols to the destination node. The feedback technique
allows the destination node to transmit the CSI or other information
back to the source node, in order to achieve gains by pre-processing
the symbols. In \cite{Mari}, the trade-off between the length of the
feedback symbols, which is related to the capacity loss, and the
transmission rate is discussed, and in \cite{Amir}, one solution for
this trade-off problem is derived. The use of limited feedback for
STC encoding has been widely discussed in the literature. In
\cite{Jabran}, the phase information is sent back for STC encoding
in order to maintain the full diversity, and the phase feedback is
employed in \cite{IIhwan} to improve the performance of the Alamouti
STBC. The limited feedback is used in \cite{George} and \cite{David}
to provide the channel information for the pre-coding of an OSTBC
scheme.

In this paper, we propose an adaptive linear receiver design
algorithm with randomized distributed space-time coding optimization
based on the MSE criterion for cooperative MIMO relaying systems
with limited feedback. We focus on how the randomized matrix affects
the DSTC during the encoding and how to optimize the linear receive
filter with the randomized matrix iteratively. It is shown that the
utilization of a randomized matrix benefits the performance of the
system compared to using traditional STC schemes. Then an adaptive
optimization algorithm is derived based on the MSE criterion subject
to constraints on the transmitted power at the relays, with the aid
of a stochastic gradient (SG) algorithm in order to release the
destination node from the high computing complexity of the
optimization process. The updated randomized matrix is transmitted
to the relay node through a feedback channel with errors, and the
influence of the imperfect feedback is discussed.

The paper is organized as follows. Section II introduces a two-hop
cooperative MIMO system with multiple relays applying the AF
strategy and the randomized DSTC scheme. In Section III the proposed
optimization algorithm for the randomized matrix is derived, and the
results of the simulations are given in Section IV and Section V
leads to the conclusion.

%\begin{figure}\label{Fig.1}
%  \includegraphics[width=3.5in]{fig5.eps}\vspace*{-1em}\\
%  \caption{Cooperative MIMO system model with $n_r$ relay nodes}\label{1}
%\end{figure}

\section{Cooperative System Model}

The communication system under consideration, shown in Fig.1, is a
cooperative communication system employing multi-antenna relay nodes
transmitting through a MIMO channel from the source node to the
destination node with feedback channels to the relay nodes. The
4-QAM modulation scheme is used in our system to generate the
transmitted symbol vector ${\boldsymbol s}[i]$ at the source node.
There are $n_r$ relay nodes with $N$ antennas for transmitting and
receiving, applying an AF cooperative strategy as well as a DSTC
scheme, between the source node and the destination node. A two-hop
communication system that broadcasts symbols from the source to
$n_r$ relay nodes as well as to the destination node in the first
phase, followed by transmitting the amplified and re-encoded symbols
from each relay node to the destination node in the next phase.
After decoding at the destination node, the information matrix for
encoding will be quantized first, and then transmitted back to each
relay node through a feedback channel with noise and interference.
The relay nodes quantize the feedback symbols and use them as a part
of the encoding matrix in the next transmission. We consider only
one user at the source node in our system that has $N$ Spatial
Multiplexing (SM)-organized data symbols contained in each packet.
The received symbols at the $k-th$ relay node and the destination
node are denoted as ${\boldsymbol r}_{{SR}_{k}}$ and ${\boldsymbol
r}_{SD}$, respectively, where $k=1,2,...,n_r$. The received symbols
${\boldsymbol r}_{{SR}_{k}}$ will be amplified before mapped into an
STC matrix. We assume that the synchronization at each node is
perfect. The received symbols at the destination node and each relay
node can be described as follows
\begin{equation}\label{2.1}
{\boldsymbol r}_{{SR}_{k}}[i]  = {\boldsymbol F}_{k}[i]{\boldsymbol s}[i] + {\boldsymbol n}_{SR_k}[i],
\end{equation}
\begin{equation}\label{2.2}
{\boldsymbol r}_{SD}[i] = {\boldsymbol H}[i]{\boldsymbol s}[i] + {\boldsymbol n}_{SD}[i],
\end{equation}
\begin{equation*}
i = 1,2,~...~,N, ~~k = 1,2,~...~ n_{r},
\end{equation*}
where the $N \times 1$ vector ${\boldsymbol n}_{{SR}_{k}}[i]$ and ${\boldsymbol n}_{SD}[i]$ denote the zero mean complex circular symmetric additive white Gaussian noise (AWGN) vector generated at each relay and the destination node with variance $\sigma^{2}$. The transmitted symbol vector ${\boldsymbol s}[i]$ contains $N$ parameters, ${\boldsymbol s}[i] = [s_{1}[i], s_{2}[i], ... , s_{N}[i]]$, which has a covariance matrix $E\big[ {\boldsymbol s}[i]{\boldsymbol s}^{H}[i]\big] = \sigma_{s}^{2}{\boldsymbol I}$, where $E[\cdot]$ stands for expected value, $(\cdot)^H$ denotes the Hermitian operator, $\sigma_s^2$ is the signal power which we assume to be equal to 1 and ${\boldsymbol I}$ is the identity matrix. ${\boldsymbol F}_k[i]$ and ${\boldsymbol H}[i]$ are the $N \times N$ channel gain matrices between the source node and the $k-th$ relay node, and between the source node and the destination node, respectively.

After processing and amplifying the received vector ${\boldsymbol
r}_{SR_k}[i]$ at the $k-th$ relay node, the signal vector
$\tilde{{\boldsymbol s}}_{SR_k}[i]={\boldsymbol
A}_{R_kD}[i]({\boldsymbol F}_{k}[i]{\boldsymbol s}[i] + {\boldsymbol
n}_{{SR}_{k}}[i])$ can be obtained and will be forwarded to the
destination node. The amplified symbols in $\tilde{{\boldsymbol
s}}_{SR_k}[i]$ will be re-encoded by an $N \times T$ DSTC scheme
${\boldsymbol M}(\tilde{{\boldsymbol s}}[i])$ and then multiplied by
an $N \times N$ randomized matrix ${\boldsymbol {\mathfrak{R}}}[i]$
in \cite{Birsen Sirkeci-Mergen}, then forwarded to the destination
node. The relationship between the $k-th$ relay and the destination
node can be described as
\begin{equation}\label{2.3}
{\boldsymbol R}_{R_{k}D}[i] = {\boldsymbol G}_k[i]{\boldsymbol {\mathfrak{R}}}[i]{\boldsymbol M}_{R_{k}D}[i] + {\boldsymbol N}_{R_{k}D}[i],
\end{equation}
\begin{equation*}
    k = 1,2, ..., n_r,
\end{equation*}
where the $N \times T$ matrix ${\boldsymbol M}_{R_{k}D}[i]$ is the DSTC matrix employed at the relay nodes whose elements are the amplified symbols in $\tilde{{\boldsymbol s}}_{SR_k}[i]$. The $N \times T$ received symbol matrix ${\boldsymbol R}_{R_{k}D}[i]$ in (\ref{2.3}) can be written as an $NT \times 1$ vector ${\boldsymbol r}_{R_{k}D}[i]$ given by
\begin{equation}\label{2.4}
{\boldsymbol r}_{R_{k}D}[i]  = \sum_{j = 1}^{N}{\boldsymbol {\mathfrak{R}}}_{{eq_k}_j}[i]{\boldsymbol G}_{{eq_k}_j}[i]\tilde{{\boldsymbol s}}_{{SR_k}_j}[i] + {\boldsymbol n}_{R_{k}D}[i],
\end{equation}
where the $NT \times N$ matrix ${\boldsymbol G}_{{eq_k}_j}[i]$ stands for the equivalent channel matrix which is the DSTC scheme ${\boldsymbol M}(\tilde{{\boldsymbol s}}[i])$ combined with the channel matrix ${\boldsymbol G}_{R_kD}[i]$ and the block diagonal $NT \times NT$ matrix ${\boldsymbol {\mathfrak{R}}}_{{eq_k}_j}[i]$ denotes the equivalent randomized matrix assigned for the $j-th$ forwarded symbol at the relay node. The $NT \times 1$ equivalent noise vector ${\boldsymbol n}_{R_kD}[i]$ generated at the destination node contains the noise parameters in ${\boldsymbol N}_{R_kD}[i]$. After rewriting ${\boldsymbol R}_{R_{k}D}[i]$ we can consider the received symbol vector at the destination node as a $N(n_r+1)$ vector with two parts, one is from the source node and another one is the superposition of the received vectors from each relay node, therefore the received symbol vector for the cooperative MIMO network we considered can be written as
\begin{equation}\label{2.5}
\begin{aligned}
{\boldsymbol r}[i]  &=
\left[\begin{array}{c} \sum_{j=1}^{N}{\boldsymbol H}_{eq_j}[i]s_j[i]  \\ \sum_{k=1}^{n_r}\sum_{j = 1}^{N}{\boldsymbol {\mathfrak{R}}}_{{eq_k}_j}[i]{\boldsymbol G}_{{eq_k}_j}[i]\tilde{s}_{{SR_k}_j}[i] \end{array} \right] \\ & ~~~~ + \left[\begin{array}{c}{\boldsymbol n}_{SD}[i] \\ {\boldsymbol n}_{RD}[i] \end{array} \right] \\
& = \sum_{j=1}^{N}{\boldsymbol D}_{D_j}[i]\tilde{\boldsymbol s}_{D_j}[i] + {\boldsymbol n}_D[i],
\end{aligned}
\end{equation}
where the $(T + 1)N \times (n_r + 1)N$ block diagonal matrix ${\boldsymbol D}_{D_j}[i]$ denotes the channel gain matrix of all the links in the network for the $j-th$ symbol in $\tilde{\boldsymbol s}_{D_j}[i]$ which contains the $N \times N$ channel coefficients matrix ${\boldsymbol H}[i]$ between the source node and the destination node, the $NT \times N$ equivalent channel matrix ${\boldsymbol G}_{{eq}_k}[i]$ for $k=1,2,...,n_r$ between each relay node and the destination node. The $(n_r + 1)N \times 1$ noise vector ${\boldsymbol n}_D[i]$ contains the received noise vector at the destination node and the amplified noise vectors from each relay node, which can be derived as an AWGN with zero mean and covariance matrix $\sigma^{2}(1+\parallel{\boldsymbol {\mathfrak{R}}}_{eq_k}[i]{\boldsymbol G}_{{eq}_k}[i]{\boldsymbol A}_{R_kD}[i]\parallel^2_F){\boldsymbol I}$, where $\parallel{\boldsymbol X}\parallel_F=\sqrt{{\rm Tr}({\boldsymbol X}^H\cdot{\boldsymbol X})}=\sqrt{{\rm Tr}({\boldsymbol X}\cdot{\boldsymbol X}^H)}$ stands for the Frobenius norm.

\section{Joint Constrained Adaptive Randomized STC Optimization and Linear MMSE Receiver Design}

As derived in the previous section, the DSTC scheme used at the
relay node will be multiplied by a randomized matrix subject to a
power constraint before being forwarded to the destination node. In
this section, we design a constrained adaptive optimization
algorithm based on an SG estimation algorithm \cite{S. Haykin} for
determining the optimal randomized matrix and the linear MMSE
receive filters.

\subsection{Linear MMSE Receiver Design with RSTC Optimization}

The linear MMSE receiver design and the optimal RSTC matrices
subject to a transmit power constraint at the relays are derived as
follows. Define the $(T+1)N \times 1$ parameter vector ${\boldsymbol
w}_j[i]$ to determine the $j-th$ symbol $s_j[i]$. From (\ref{2.5})
we propose the MSE based optimization with a power constraint at the
destination node as
\begin{equation*}
    [{\boldsymbol w}_j[i],{\boldsymbol {\mathfrak{R}}}_{{eq_k}_j}[i]] = \arg\min_{{\boldsymbol w}_j[i], {\boldsymbol {\mathfrak{R}}}_{{eq_k}_j}[i]} E\left[\|s_j[i]-{\boldsymbol w}_j^H[i]{\boldsymbol r}[i]\|^2\right],
\end{equation*}
subject to
\begin{equation*}
    \sum_{j=1}^{N}\rm{trace}({\boldsymbol {\mathfrak{R}}}_{{eq_k}_j}[i]{\boldsymbol {\mathfrak{R}}}_{{eq_k}_j}^H[i])\leq \rm{P_R},
\end{equation*}
where ${\boldsymbol r}[i]$ denotes the received symbol vector at the
destination node which contains the randomized STC matrix with the
power constraint of $P_R$. If we only consider the received symbols
from the relay node, the received symbol vector at the destination
node can be derived as
\begin{equation}\label{4.1}
\begin{aligned}
    {\boldsymbol r}[i] &=\sum_{k=1}^{n_r}\sum_{j = 1}^{N}{\boldsymbol {\mathfrak{R}}}_{{eq_k}_j}[i]{\boldsymbol G}_{{eq_k}_j}[i]\tilde{{\boldsymbol s}}_{{SR_k}_j}[i] + {\boldsymbol n}_D[i]\\
    &=\sum_{k=1}^{n_r}\sum_{j = 1}^{N}{\boldsymbol {\mathfrak{R}}}_{{eq_k}_j}[i]{\boldsymbol G}_{{eq_k}_j}[i]{\boldsymbol A}_j[i]{\boldsymbol F}_j[i]s_j[i]\\ &~~~+\sum_{k=1}^{n_r}\sum_{j = 1}^{N}{\boldsymbol {\mathfrak{R}}}_{{eq_k}_j}[i]{\boldsymbol G}_{{eq_k}_j}[i]{\boldsymbol A}_j[i]n_{SR_j}[i]\\
    &~~~+{\boldsymbol n}_{RD}[i]\\
    &=\sum_{k=1}^{n_r}\sum_{j = 1}^{N}{\boldsymbol {\mathfrak{R}}}_{{eq_k}_j}[i]{\boldsymbol C}_{k_j}[i]s_j[i]+{\boldsymbol n}_{D_{eq}}[i],
\end{aligned}
\end{equation}
where ${\boldsymbol C}_{k_j}[i]$ is an $NT \times N$ matrix that
contains all the complex channel gains and the amplified matrix
assigned to the received symbol $s_j[i]$ at the relay node, and the
noise vector ${\boldsymbol n}_{D_{eq}}[i]$ is a Gaussian noise
vector with zero mean and variance
$\sigma^{2}(1+\sum_{k=1}^{n_r}\parallel{\boldsymbol
{\mathfrak{R}}}_{{eq_k}_j}[i]{\boldsymbol
G}_{{eq_k}_j}[i]{\boldsymbol A}_j[i]\parallel^2_F)$. Therefore, we
can rewrite the MSE cost function as in (\ref{4.2}).
\begin{table*}
\begin{equation}\label{4.2}
    [{\boldsymbol w}_j[i],{{\boldsymbol {\mathfrak{R}}}_{{eq_k}_j}}[i]] = \arg\min_{{\boldsymbol w}_j[i], {{\boldsymbol {\mathfrak{R}}}_{{eq_k}_j}}[i]} E\left[\|s_j[i]-{\boldsymbol w}_j^H[i](\sum_{k=1}^{n_r}\sum_{j = 1}^{N}{{\boldsymbol {\mathfrak{R}}}_{{eq_k}_j}}[i]{\boldsymbol C}_{k_j}[i]s_j[i]+{\boldsymbol n}_D[i])\|^2\right],~~~s.t.~~~ \sum_{j=1}^{N}\rm{trace}({\boldsymbol {\mathfrak{R}}}_{{eq_k}_j}[i]{\boldsymbol {\mathfrak{R}}}_{{eq_k}_j}^H[i])\leq \rm{P_R}.
\end{equation}
\rule{18cm}{1pt}
\end{table*}

Since ${\boldsymbol w}_j[i]$ can be optimized by expanding the
righthand side of (\ref{4.2}) and taking the gradient with respect
to ${\boldsymbol w}_j^*[i]$ and equating the terms to zero, we can
obtain the $j-th$ MMSE receive filter
\begin{equation}\label{4.3}
    {\boldsymbol w}_j[i]=\left(E\left[{\boldsymbol r}[i]{\boldsymbol r}^H[i]\right]\right)^{-1}E\left[{\boldsymbol r}[i]s_j^H[i]\right],
\end{equation}
where $E\left[{\boldsymbol r}[i]{\boldsymbol r}^H[i]\right]$ denotes
the auto-correlation matrix and $E\left[{\boldsymbol
r}[i]s_j^H[i]\right]$ stands for the cross-correlation matrix. By
optimizing the randomized matrix ${\boldsymbol
{\mathfrak{R}}}_{{eq_k}_j}[i]$ for each symbol at each relay node,
we can first define a vector $\tilde{\boldsymbol
r}_{{eq_k}_j}={\boldsymbol C}_{k_j}[i]s_j[i]+{\boldsymbol
C}_{k_j}[i]n_{SR_j}$, where the parameter $n_{SR_j}$ denotes the
$j-th$ symbol in the noise vector ${\boldsymbol n}_{SR}$, then the
randomized matrix can be calculated by taking the gradient with
respect to ${\boldsymbol {\mathfrak{R}}}_{{eq_k}_j}^*[i]$ and
equating the term to zero, resulting in
\begin{equation}\label{4.4}
\begin{aligned}
    {\boldsymbol {\mathfrak{R}}}_{{eq_k}_j}[i]=&
\left({\boldsymbol w}_j^H[i](E\left[\tilde{\boldsymbol
r}[i]\tilde{\boldsymbol r}^H[i]\right]){\boldsymbol w}_j[i] +
\lambda {\boldsymbol I} \right)^{-1}\\ &~~~~
E\left[\tilde{\boldsymbol r}^H[i]s_j[i]\right]{\boldsymbol w}_j[i],
\end{aligned}
\end{equation}
where $E\left[\tilde{\boldsymbol r}[i]\tilde{\boldsymbol
r}^H[i]\right]$ denotes the auto-correlation matrix of the
equivalent space-time coded received symbol vector without the
randomized matrix at the relay node, and $E\left[\tilde{\boldsymbol
r}^H[i]s_j[i]\right]$ denotes the cross-correlation matrix. The
power constraint can be achieved by multiplying the quotient of
$P_R$ and the trace of the updated randomized matrix. The expression
in (\ref{4.4}) does not provide a closed-form solution of the
randomized STC matrix ${\boldsymbol {\mathfrak{R}}}_{{eq_k}_j}[i]$
assigned for the $j-th$ received symbol at the $k-th$ relay node
because it requires the adjustment of the Lagrange multiplier
$\lambda$. This parameter needs to be adjusted in order to enforce
the power constraint. Moreover, the expression in (\ref{4.4} also
requires an inversion calculation with a high computational
complexity. With the increase of the number of antennas employed at
each node or employing more complicated STC encoders at the relay
nodes, the complexity increases exponentially according to the
matrix size in (\ref{4.4}).

\subsection{Adaptive Linear MMSE Receiver Design with Randomized Matrix Optimization Algorithm}

In order to reduce the computational complexity of the proposed
design and compute the required parameters, an adaptive linear
receiver design with randomized matrix optimization (ALRRMO)
algorithm is proposed. We resort to a strategy that initially drops
the power constraint, obtain the necessary recursions and then
enforce the constraint with a normalization step. We define the
Lagrangian of the constrained MSE minimization problem in
(\ref{4.2}) as
\begin{equation}
\begin{split}
{\mathcal L}({\boldsymbol w}_j[i],{\boldsymbol
{\mathfrak{R}}}_{{eq_k}_j}[i]]) & = E\left[\|s_j[i]-{\boldsymbol
w}_j^H[i]{\boldsymbol r}[i]\|^2\right] \\ & \quad + \left(
\sum_{j=1}^{N}\rm{trace}({\boldsymbol
{\mathfrak{R}}}_{{eq_k}_j}[i]{\boldsymbol
{\mathfrak{R}}}_{{eq_k}_j}^H[i]) - \rm{P_R} \right)\lambda,
\end{split}
\end{equation}
A simple adaptive algorithm for determining the linear receive
filters and the randomized matrices can be achieved by taking the
instantaneous gradient term of (\ref{4.2}) with respect to
${\boldsymbol w}_j^*[i]$ and with respect to ${{\boldsymbol
{\mathfrak{R}}}_{{eq_k}_j}}^*[i]$, respectively, which are
\begin{equation}\label{4.41}
\begin{aligned}
     \nabla {\rm L}_{{\boldsymbol w}_j^*[i]} & = \nabla E\left[\|s_j[i]-{\boldsymbol w}_j^H[i]{\boldsymbol r}[i]\|^2\right]_{{\boldsymbol w}_j^*[i]}\\ & = -(s_j[i]-{\boldsymbol w}_j^H[i]{\boldsymbol r}[i])^H{\boldsymbol r}[i] = -e_j^*[i]{\boldsymbol r}[i],\\
     \nabla {\rm L}_{{\boldsymbol {\mathfrak{R}}}_{{eq_k}_j}^*[i]} &= \nabla E\left[\|s_j[i]-{\boldsymbol w}_j^H[i]{\boldsymbol r}[i]\|^2\right]_{{{\boldsymbol {\mathfrak{R}}}^*_{{eq_k}_j}}[i]} \\
    &= -e_j[i]s_j^H[i]{\boldsymbol C}_{k_j}^H[i]{\boldsymbol w}_j[i],
\end{aligned}
\end{equation}
where $e_j[i]$ stands for the $j-th$ detected error. After we obtain
(\ref{4.41}) the proposed algorithm is obtained by introducing a
step size into the recursions. The proposed algorithm is given by
\begin{equation}\label{4.7}
\begin{aligned}
     {\boldsymbol w}_j[i+1] & = {\boldsymbol w}_j[i] + \beta (e_j^*[i]{\boldsymbol r}[i]),\\
     {\boldsymbol {\mathfrak{R}}}_{{eq_k}_j}[i+1] &={\boldsymbol {\mathfrak{R}}}_{{eq_k}_j}[i] +\mu (e_j[i]s_j^H[i]{\boldsymbol C}_{k_j}^H[i]{\boldsymbol w}_j[i]),\\
     {\boldsymbol {\mathfrak{R}}}_{{eq_k}_j}[i+1] &=\frac{\sqrt{\rm{P_R}}{\boldsymbol {\mathfrak{R}}}_{{eq_k}_j}[i+1]}{\sqrt{\sum_{j=1}^{N}\rm{trace}({\boldsymbol {\mathfrak{R}}}_{{eq_k}_j}[i+1]{\boldsymbol {\mathfrak{R}}}_{{eq_k}_j}^H[i+1])}},
\end{aligned}
\end{equation}
where $\beta$ and $\mu$ denote the step sizes for the recursions for
the estimation of the linear MMSE receive filter and the randomized
matrix in the RSTC scheme, respectively. The last equation in
(\ref{4.7}) stands for the normalization of the randomized matrix
after the iteration. According to (\ref{4.7}), the desired vector
and the matrix depends on each other, so that the algorithm in
\cite{RC De Lamare2008} can be used to determine the linear MMSE
receive filter and the randomized matrix iteratively, and the design
can be achieved. The complexity for calculating the optimal
${\boldsymbol w}_j[i]$ and ${\boldsymbol
{\mathfrak{R}}}_{{eq_k}_j}[i]$ is ${\rm O}(N(T+1))$ and ${\rm
O}(N^2T^2)$, respectively, which is much less than $O(2N^3(T+1)^3)$
and $O(2N^4T^4)$ by using (\ref{4.3}) and (\ref{4.4}). As mentioned
in Section I, the randomized matrix will be sent back to the relay
nodes via a feedback channel which requires quantization as will be
shown in the simulations.

\section{Simulations}

The simulation results are shown here to assess the proposed scheme
and algorithm. The system we considered is an AF cooperative MIMO
system with the Alamouti STBC scheme using QPSK modulation in
quasi-static block fading channel with AWGN, as derived in Section
II. The bit error rate (BER) performance of the proposed adaptive
linear receiver design with RSTC optimization algorithm is assessed,
and the influence of the imperfect feedback channels are considered
in the simulations. The system employs 1 relay node and each node in
the system has 2 antennas. In the simulation, we define both the
symbol power at the source node and the noise variance $\sigma^{2}$
for each link to be equal to 1. The RSTC scheme is designed by
multiplying the 2 $\times$ 2 Alamouti STBC \cite{Alamouti} by a
randomized matrix with each element generated using $e^{j\theta}$
where $\theta$ is uniformly distributed in $[0,2\pi)$.

The proposed ALRRMO algorithm is compared with the SM scheme and the
traditional RSTC algorithm using the distributed-Alamouti
(D-Alamouti) STBC scheme in \cite{RC De Lamare} with $n_r = 1$ relay
nodes in Fig. 2. The number of antennas $N=2$ at each node and the
effect of the direct link is considered. The results illustrate that
without the direct link, by making use of the STC or the RSTC
technique, a significant performance improvement can be achieved
compared to the spatial multiplexing system.  The RSTC algorithm
outperforms the STC-AF system, while the ALRRMO algorithm can
improve the performance by about 3dB as compared to the RSTC
algorithm. With the consideration of the direct link, the results
indicate that the cooperative diversity order can be increased, and
using the ALRRMO algorithm achieves an improved performance with
$2$dB of gain as compared to employing the RSTC algorithm and $3$dB
of gain as compared to employing the traditional STC-AF algorithm.

%\begin{figure}
%\begin{center}
%\def\epsfsize#1#2{0.95\columnwidth}
%\epsfbox{fig1.eps}\vspace*{-1em} \caption{BER performance v.s.
%$E_b/N_0$ for ALRRMO Algorithm with and without the Direct
%Link}\label{2}
%\end{center}
%\end{figure}

The simulation results shown in Fig. 3 illustrate the impact of the
feedback channel for the ALRRMO algorithm. As mentioned in Section
I, the optimal randomized matrix will be sent back to each relay
node through a feedback channel. The quantization and feedback
errors are not considered in the simulation results in Fig. 2, so
the optimal randomized matrix is perfectly known at the relay node
after the ALRRMO algorithm; while in Fig. 3, it indicates that the
performance of the proposed algorithm will be affected by the
accuracy of the feedback information. In the simulation, we use 4
bits to quantize the real part and the imaginary part of each
element of the randomized matrix ${\boldsymbol
{\mathfrak{R}}}_{{eq_k}_j}[i]$, and the feedback channel is a binary
symmetric channel. As we can see from Fig. 3, by decreasing the
error probabilities for the feedback channel with fixed quantization
bits, the BER performance approaches the performance with the
perfect feedback, and by making use of 4 quantization bits for the
real and imaginary part of each parameter in the randomized matrix,
the performance of the ALRRMO algorithm is about 1dB worse with
feedback error probability of $10^{-3}$.

%\begin{figure}
%\begin{center}
%\def\epsfsize#1#2{0.95\columnwidth}
%\epsfbox{fig2.eps}\vspace*{-1em} \caption{BER performance v.s.
%number of samples for ALRRMO algorithm with perfect and imperfect
%feedback links, quantization bits = 4}\label{3}
%\end{center}
%\end{figure}

In Fig. 4 and Fig. 5, the influence of different feedback error
probabilities with various quantization bits are employed to test
the performance of the ALRRMO algorithm. In Fig. 4, the BER
performance with a perfect feedback channel is given as a lower
bound with $SNR=15dB$ and $SNR=30dB$, respectively. The error
probability is fixed in $P_e=10^{-3}$. With the increase of the
number of bits we employed in the quantization, the BER curves will
approach the result with perfect feedback due to a more accurate
estimation with the cost of computing complexity increase. With more
quantization bits, the ALRRMO algorithm can achieve a performance as
good as that with perfect feedback. However, it is worth to mention
that the BER decreases slightly when we increase the number of
quantization bits to 5 and 6. If we fix the number of quantization
bits to 4, the BER performance gets worse with the increase of the
feedback error probability as depicted in Fig.5. By increasing the
feedback error probability, the BER curves become gradually worse.
Thus, there is a trade-off between the feedback error probability
and the number of quantization bits. As indicated by the simulation
results in Fig. 3 to Fig. 5, the 4-bit quantization is an
appropriate choice.

%\begin{figure}
%\begin{center}
%\def\epsfsize#1#2{1\columnwidth}
%\epsfbox{fig3.eps}\vspace*{-1em} \caption{BER performance v.s.
%different number of bits for feedback quantization,
%$P_e=10^{-3}$}\label{4}
%\end{center}
%\end{figure}
%
%\begin{figure}
%\begin{center}
%\def\epsfsize#1#2{1\columnwidth}
%\epsfbox{fig4.eps}\vspace*{-1em} \caption{BER performance v.s.
%different error probabilities for feedback channel, quantization
%bits = 4}\label{5}
%\end{center}
%\end{figure}

\section{Conclusion}

We have proposed an adaptive linear receiver filter design with
randomized matrix optimization (ALRRMO) algorithm for the randomized
DSTC in a cooperative system. A joint iterative estimation algorithm
for computing the receive filters and the randomized matrix has been
derived. The effect of the limited feedback and feedback errors are
considered in the simulation. The simulation results illustrate the
advantage of the proposed ALRRMO algorithm by comparing it with the
cooperative network employing the traditional DSTC scheme and the
fixed randomized STC scheme. The proposed algorithm can be used with
different distributed STC schemes using the AF strategy and can also
be extended to the DF cooperation protocol.

%\section{Outlook}

%In the full paper, we will present a complete description of the algorithm and further results with more discussions and a complete set of references.

\ifCLASSOPTIONcaptionsoff
  \newpage
\fi

\bibliographystyle{IEEEtran}

\end{document}